\documentstyle[12pt,epsfig]{article}
\title{\hspace*{10cm} {\large Budker INP 98-33 \\
}
\vspace{0.5cm} 
 {\bf Electron-photon interactions in high energy beam production and 
cooling}\thanks{Talk at ICFA Workshop {\it Quantum Aspects of Beam Physics},
Monterey, CA, USA, January 4-9, 1998. To be published by World
Scientific.}}
\author{Valery Telnov\thanks{Email: telnov@inp.nsk.su} \\
{\small \it Institute of Nuclear Physics,
630090, Novosibirsk, Russia}} 
\date{}
\begin{document}
\newcommand{\M}{\mbox{m}}
\newcommand{\n}{\mbox{$n_f$}}
\newcommand{\EP}{\mbox{e$^+$}}
\newcommand{\EM}{\mbox{e$^-$}}
\newcommand{\EPEM}{\mbox{e$^+$e$^-$}}
\newcommand{\EMEM}{\mbox{e$^-$e$^-$}}
\newcommand{\GG}{\mbox{$\gamma\gamma$}}
\newcommand{\GE}{\mbox{$\gamma$e}}
\newcommand{\GP}{\mbox{$\gamma$e$^+$}}
\newcommand{\TEV}{\mbox{TeV}}
\newcommand{\GEV}{\mbox{GeV}}
\newcommand{\LGG}{\mbox{$L_{\gamma\gamma}$}}
\newcommand{\LEE}{\mbox{$L_{ee}$}}
\newcommand{\WGG}{\mbox{$W_{\gamma\gamma}$}}
\newcommand{\EV}{\mbox{eV}}
\newcommand{\CM}{\mbox{cm}}
\newcommand{\MM}{\mbox{mm}}
\newcommand{\NM}{\mbox{nm}}
\newcommand{\MKM}{\mbox{$\mu$m}}
\newcommand{\SEC}{\mbox{s}}
\newcommand{\CMS}{\mbox{cm$^{-2}$s$^{-1}$}}
\newcommand{\MRAD}{\mbox{mrad}}
\newcommand{\IND}{\hspace*{\parindent}}
\newcommand{\E}{\mbox{$\epsilon$}}
\newcommand{\EN}{\mbox{$\epsilon_n$}}
\newcommand{\EI}{\mbox{$\epsilon_i$}}
\newcommand{\ENI}{\mbox{$\epsilon_{ni}$}}
\newcommand{\ENX}{\mbox{$\epsilon_{nx}$}}
\newcommand{\ENY}{\mbox{$\epsilon_{ny}$}}
\newcommand{\EX}{\mbox{$\epsilon_x$}}
\newcommand{\EY}{\mbox{$\epsilon_y$}}
\newcommand{\BI}{\mbox{$\beta_i$}}
\newcommand{\BX}{\mbox{$\beta_x$}}
\newcommand{\BY}{\mbox{$\beta_y$}}
\newcommand{\SX}{\mbox{$\sigma_x$}}
\newcommand{\SY}{\mbox{$\sigma_y$}}
\newcommand{\SZ}{\mbox{$\sigma_z$}}
\newcommand{\SI}{\mbox{$\sigma_i$}}
\newcommand{\SIP}{\mbox{$\sigma_i^{\prime}$}}
\maketitle
\begin{abstract} 
\IND\ In this review we consider three important
applications of lasers in high energy physics: \GG, \GE\ colliders,
laser cooling, positron production. These topics are actual now due to
plans of construction linear \EPEM, \EMEM, \GG, \GE\ colliders with
energies 0.3--1 TeV. High energy photons for \GG, \GE\ collisions
can be obtained using laser backscattering. These types of collisions
considerably increase physics potential of linear colliders.  Very low
emittance of electron beams required for achieving ultimate \GG\
luminosity can be obtained using a laser cooling of electron beams.
Combining a laser-electron Compton scattering with subsequent
conversion of these photons to \EPEM\ pairs on the
target (it can be a laser target) one can get a nice source of
polarized positron.  In this paper, we briefly considered these
subjects with emphasis on underlying physics of photon-electron
interactions.
\end{abstract}
\section{Introduction}
\IND\  By ``Electron-photon interactions ...'' in the title of this
paper we imply the  electron -- laser beam interactions. Fantastic
progress in laser technique makes it possible now to consider
seriously many various applications of lasers in particle physics. In
this talk I will consider only three subjects connected with
development of high energy linear colliders:

\vspace{0.3cm}

\noindent$\bullet$ High energy gamma-gamma, gamma-electron colliders.\\ 
$\bullet$ Laser cooling of electron beams.\\ 
$\bullet$ Positron production for \EPEM\ colliders.
\section{Photon colliders}
\subsection{Goals and principles of photon colliders}
\IND\ It is very likely that linear colliders with the c.m.s energies of
0.3--2 TeV will be built in about ten years from
now~\cite{LOEW}. Besides \EPEM collisions, linear colliders give us an
unique possibility to study \GG\ and \GE\ interactions at energies
and luminosities comparable to those in \EPEM\
collisions~\cite{GKST81}-\cite{BERK}.

\begin{figure}[!hbt]
\hspace*{1cm}\begin{minipage}[b]{0.45\linewidth}
\centering
\vspace*{-0.0cm} 
\hspace*{.0cm} \epsfig{file=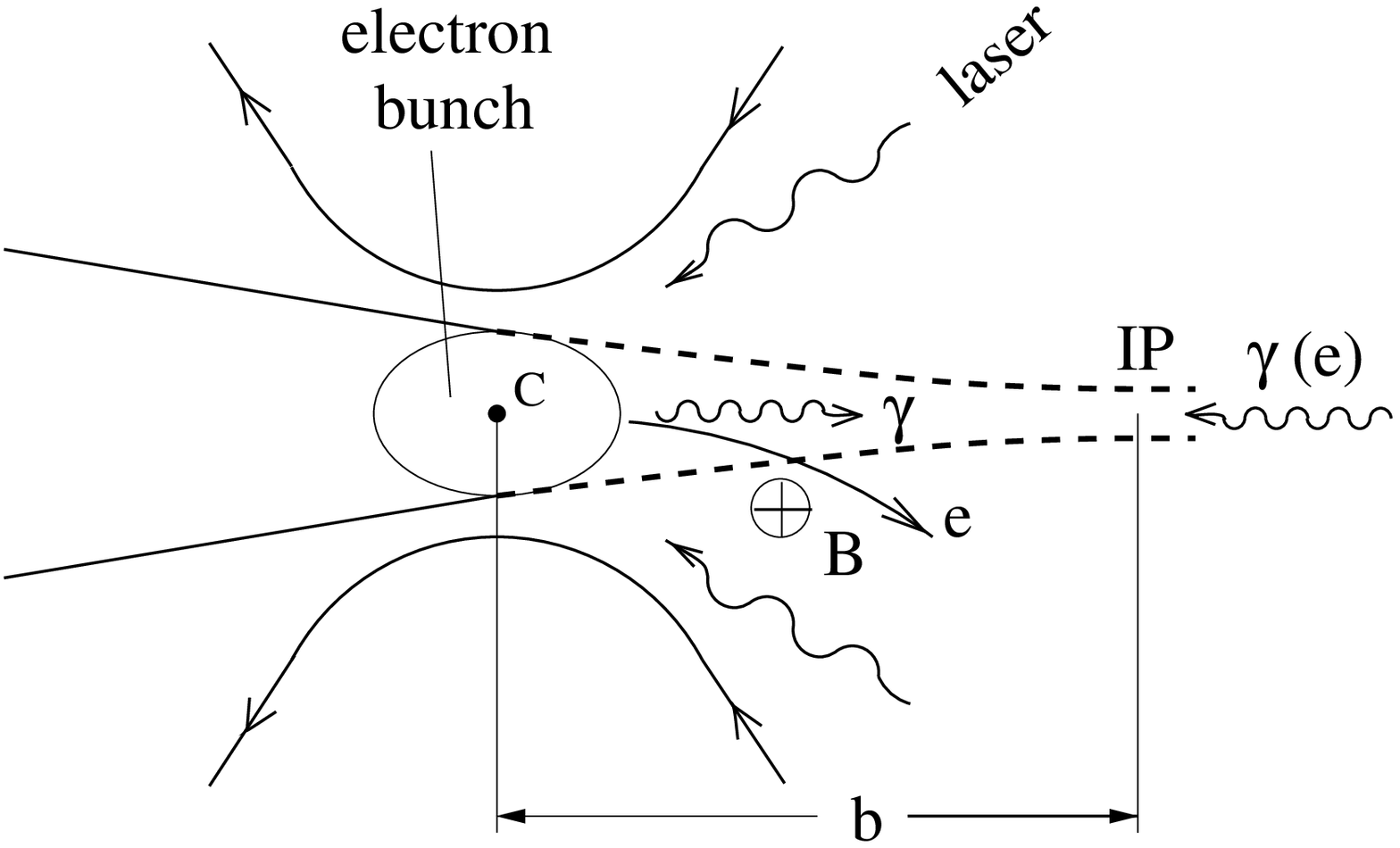,height=6.5cm,angle=-90} 
\vspace*{-0.3cm}
\caption{Scheme of  \GG; \GE\ collider.}
\vspace*{0.0cm}
\label{ris1}
\end{minipage}%
\hspace*{0.3cm} \begin{minipage}[b]{0.45\linewidth}
\centering
\vspace*{-0.0cm} 
%\hspace*{0cm} \includegraphics[width=1.5in,angle=-90]{fig9.eps}
\hspace*{0.5cm} \epsfig{file=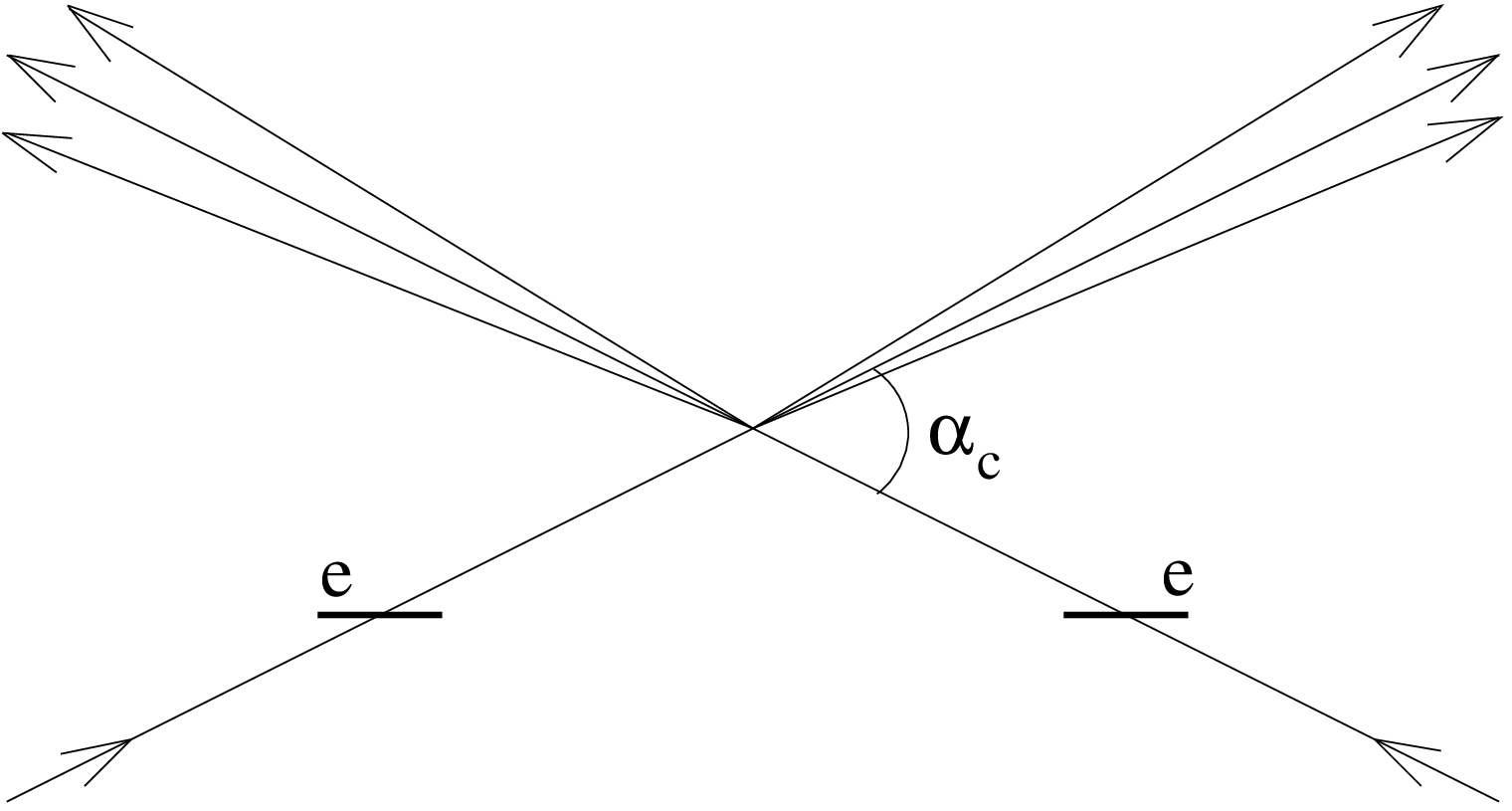,width=0.8in,angle=-90}
\vspace*{0.9cm}
\caption{Crab-crossing scheme}
\vspace*{-0.cm}
\label{fig9}
\end{minipage}
\end{figure}

The basic scheme of a photon collider is shown in
fig.\ref{ris1}. 
Two electron beams after the final focus system are traveling toward
the interaction point (IP) and at a distance of about 0.1--1 cm from
the IP collide with the focused laser beams. Photons after the Compton
scattering have the energies comparable with the energy of the initial
electrons and follow their direction (to the IP) with some small
additional angular spread of the order $1/\gamma $.  With reasonable
laser parameters one can ``convert'' most of the electrons into high
energy photons.  The luminosity of \GG, \GE\ collisions will be of the
same order of magnitude as the ``geometric'' luminosity of the basic
$ee$ beams. Luminosity distributions in \GG\ collisions have the
characteristic peaks near the maximum invariant masses with a typical
width about 10 \% (and a few times smaller in \GE\ collisions).  High
energy photons  can have various polarizations, which is very
advantageous for experiments.

     The physics at high energy \GG, \GE\ colliders is very rich and
no less interesting than that in \EPEM\ or pp collisions~\cite{BERK,TELee}.
Some examples are given below:
   
1. Some phenomena can be better studied at photon colliders than with
pp or \EPEM\ collisions, for example, the measurement of the
two-photon decay width of the Higgs boson.  Some Higgs decay
modes and its mass can be measured at \GG\ colliders more precisely than in
\EPEM\ collisions due to larger production cross sections and very
sharp edge of the luminosity spectrum.

2. Cross sections for production of charged scalar, lepton and top
  pairs in \GG\ collisions are larger than those in \EPEM\ collisions
  approximately by a factor of 5; for WW production, this factor is
  even larger, about 10--20.

3. In \GE\ collisions, charged supersymmetric particles with masses
  higher than in \EPEM\ collisions can be produced (a heavy charged
  particle plus a light neutral), \GG\ collisions also provide higher
  accessible masses for particles which are produced as a single
  resonance in \GG\ collisions (such as the Higgs boson).

These examples together with the fact that the luminosity in \GG\
collisions is potentially higher than that in \EPEM\ collisions (due
to difference in collisions effects) are very strong arguments in
favor of photon colliders.
This option has been included now into the Conceptual Design
Reports of NLC~\cite{NLC}, TESLA--SBLC~\cite{TESLA}, and
JLC~\cite{JLC} linear colliders. All these projects foresee the second
interaction regions for \GG, \GE\ collisions. 
%
%\begin{figure}[!htb]
%\vspace{-0.5cm}

%\hspace*{0cm}\begin{minipage}[b]{0.45\linewidth}
%\centering
%\vspace*{-0.cm} 
%\hspace*{0.cm} \epsfig{file=cross.eps,width=9.cm,angle=0} 
%\vspace{-0.4cm}\
%%\label{cross}
%\end{minipage}%
%\hspace*{1.1cm} \begin{minipage}[b]{0.45\linewidth}
%\centering

%\caption{Cross sections for the Standard model Higgs in \GG\ and
% \EPEM\ collisions.}
%\vspace{5cm}     
%\label{cross}
%\end{minipage}
%\end{figure} 

%\vspace{-1.5cm}

%\begin{figure}[thb]
%\centering
%\vspace*{-1.4cm}
%\epsfig{file=fig16.eps,width=6.5in}
%\vspace*{-1.2cm}
%\caption{ Comparison of  cross sections for charged pair production
%in \EPEM\ and \GG\ collisions. The cross section  $\sigma
%=(\pi\alpha^2/M^2)f(x)$, P=S (scalars), F (fermions),
%W (W-bosons); M is particle mass, $x=W_{p\bar{p}}^2/4M^2$. The functions
%$f(x)$ are shown.}

%\vspace{-0.3cm}

%\label{fig16}
%\end{figure}
%
\subsection{Effects in the conversion region}{\label{conv}}
   In the conversion region, the main process is the Compton scattering
of laser photons on high energy electrons.  Besides, there are
several other important effects.

  1) The first one is \EPEM\ pair creation in collisions of  high
energy photons with  laser photons. Above the threshold, the process
$\gamma\gamma_{L} \to$ \EPEM\ has the cross section even
larger than that of the Compton scattering. The optimum laser wave
length for photon colliders corresponds to this threshold.

2) Nonlinear QED effects (multiphoton processes) in interactions of
electrons with laser wave. Strong laser field leads to change of the
energy spectra of the scattered photons, makes possible 
\EPEM\ pair creation at electron beam
energies lower than that in the case of weak laser field.

3) A laser target is a medium with some
anisotropic refraction index. As a result, a polarization of
high energy photons after the Compton scattering can be noticeably changed
during its propagation through a laser target. 

4) Polarization of electrons is
changed obviously when they scatter. But it is less obvious that the
polarization is changed  for  electrons which passed a 
laser target without Compton scattering. 
\subsubsection{Compton scattering (linear)}
\IND\ Basic formulae for the Compton scattering in a form convenient
for the case of our interest are well known~\cite{GKST83,GKST84}.

\begin{figure}[htb]
\begin{minipage}[b]{0.45\linewidth}
\centering
\vspace*{-0.8cm} 
\hspace*{-1cm} \epsfig{file=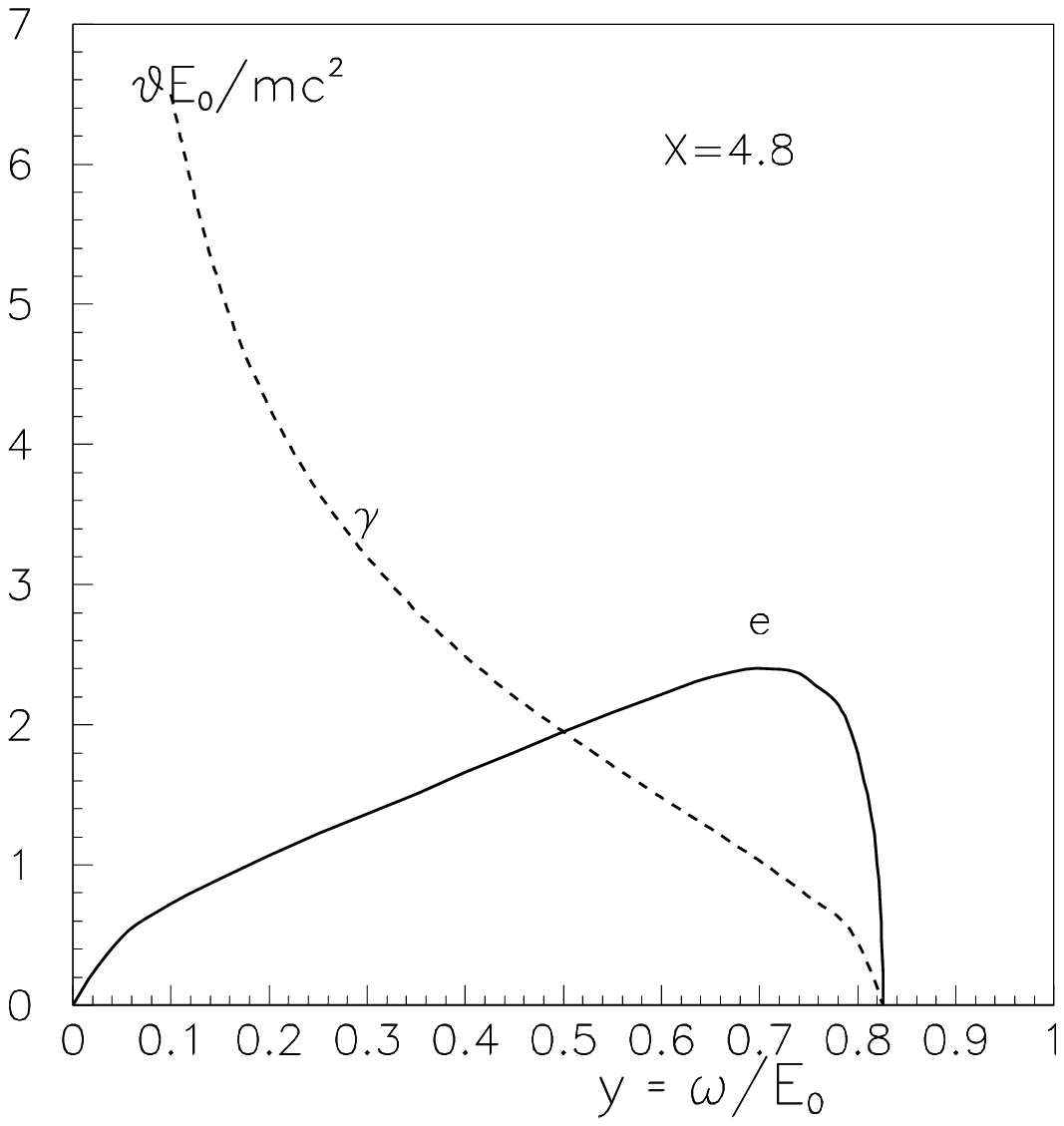, width=7.3cm}
\vspace*{-1.2cm}
\caption{Electron and photon  scattering angles vs 
photon energy for $x$=4.8  \vspace*{0.0cm} }
\vspace*{-0.0cm}
\label{fig3}
\end{minipage}%
\hspace*{0.6cm}\begin{minipage}[b]{0.45\linewidth}
\centering
\vspace{-0.8cm} 
\hspace*{-1.cm} \epsfig{file=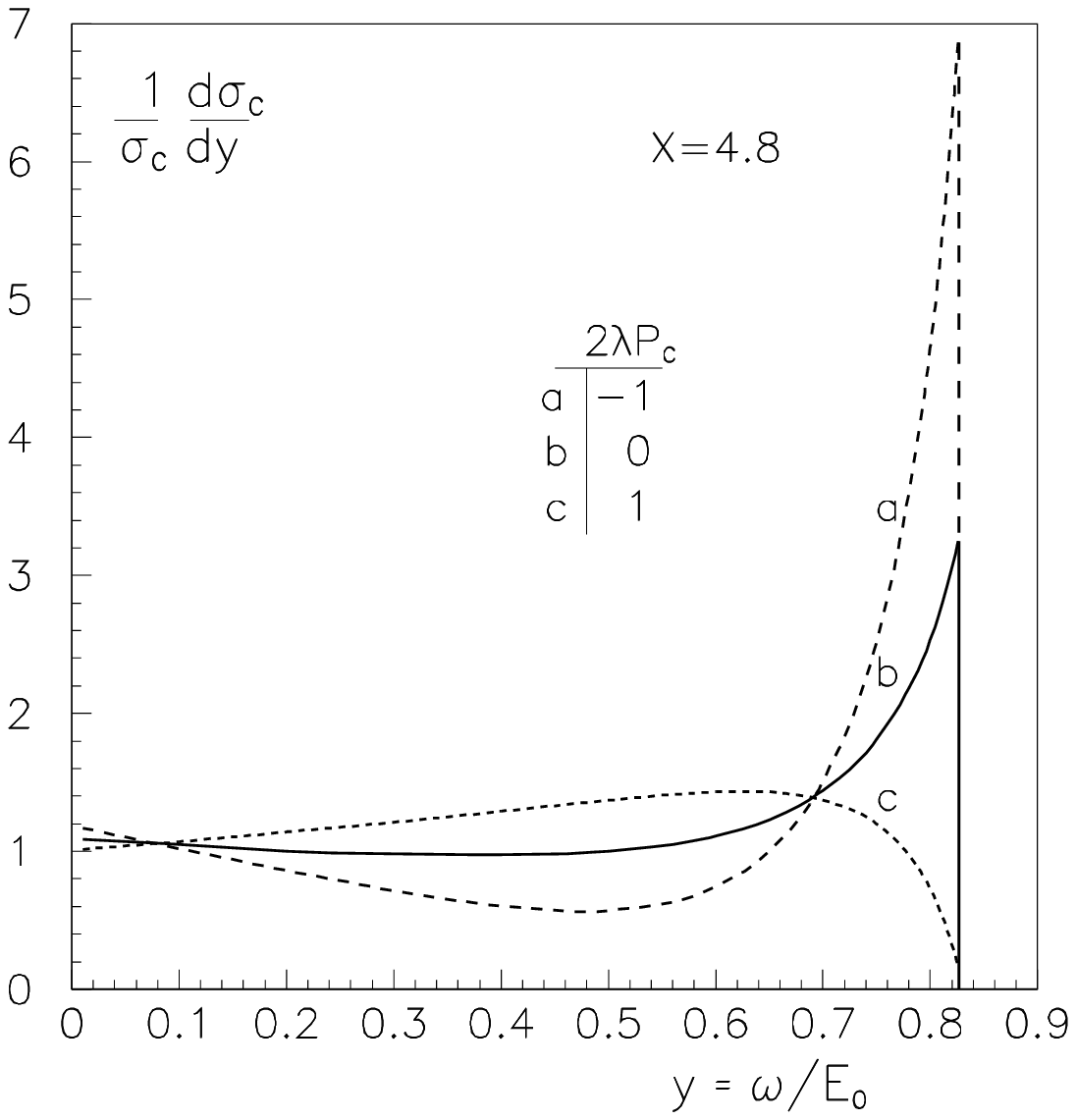, width=7.3cm}
\vspace*{-1.2cm}
\caption{Spectrum of the Compton scattered photons for different
polarizations of laser and electron beams}
\label{fig4}
\vspace*{-0.0cm}
\end{minipage}
\end{figure}

    In the conversion region a photon with an  energy $\omega_0$  is
scattered on an  electron  with  an  energy $ E_0$   at  a  small
collision angle $\alpha_0$ (almost head-on). The energy of the scattered
photon $\omega$ depends on its angle $\vartheta$ with respect to the
motion of the incident electron as follows:
\begin{equation}
\omega = \frac{\omega_m}{1+(\vartheta/\vartheta_0)^2};
\mbox{\hspace{1cm}} \omega_m=\frac{x}{x+1}E_0; \mbox{\hspace{1cm}}
\vartheta_0= \frac{mc^2}{E_0} \sqrt{x+1},
\end{equation}
$$x=\frac{4E_0 \omega_0 \cos^2 \alpha_0/2}{m^2c^4}
 \simeq 15.3\left[\frac{E_0}{TeV}\right]
\left[\frac{\omega_0}{eV}\right],$$
$\omega_m$  is the maximum photon energy,

  For example: $E_0$ =300\,\, GeV, $\omega_0 =1.17$ eV (neodymium
glass laser) $\Rightarrow$ x=5.37 and $\omega/E_0 = 0.84$.  The value
$x=4.8$ is the threshold for \EPEM\ production (see next section).
The angles of scattered particles as functions of the photon energy
for  $x=4.8$ are displayed  in fig.\ref{fig3}.

 The energy spectrum of the scattered photons (without multiple scattering)
is given by the Compton cross section
\begin{equation}
\frac{1}{\sigma_c} \frac{d\sigma_c}{dy} =
\frac{2\sigma_0}{x\sigma_c}\left[\frac{1}{1-y}+1-y-4r(1-r)+
2\lambda_eP_crx(1-2r)(2-y)\right];
\label{difspec}
\end{equation}
$$y= \leq y_m = \frac{x}{x+1}; \;\; r=\frac{y}{x(1-y)} \leq 1; \;\;
\sigma_0=\pi\left(\frac{e^2}{mc^2}\right)^2=2.5\times 10^{-25} \CM^2,$$
$\lambda_e (|\lambda_e| \leq 1/2)$ is the electron longitudinal
polarization and $P_c$ is the mean helicity of laser photons.
 The total Compton cross section is
\begin{equation}
\sigma_c = \sigma_c^{np}+2\lambda_e P_c \sigma_1 ,
\end{equation}
$$\sigma_c^{np}=\frac{2\sigma_0}{x}\left[\left(1-\frac{4}{x}-
\frac{8}{x^2}\right)\ln(x+1)+\frac{1}{2}+\frac{8}{x}-\frac{1}{2(x+1)^2}
\right],$$
$$\sigma_1=\frac{2\sigma_0}{x}\left[\left(1+\frac{2}{x}
\right)\ln(x+1)-\frac{5}{2}+\frac{1}{x+1}-\frac{1}{2(x+1)^2}
\right].$$

    In the region $x$ = 1--10 the ratio $|\sigma_1/ \sigma_c| < 0.2$ ,
i.e. the total cross section only slightly depends on the
polarization. However, the energy spectrum does essentially depend on
the value of $2\lambda P_c$. At $2\lambda P_c =-1$ and $x = 4.8$ the
relative number of hard photons doubles (see fig.\ref{fig4}),
improving essentially the monochromaticity of the photon beam. 

\begin{figure}[htb]
\vspace*{-1.3cm} 
\begin{minipage}[b]{0.45\linewidth}
\centering
\vspace*{-0.0cm} 
\hspace*{-0.cm} \epsfig{file=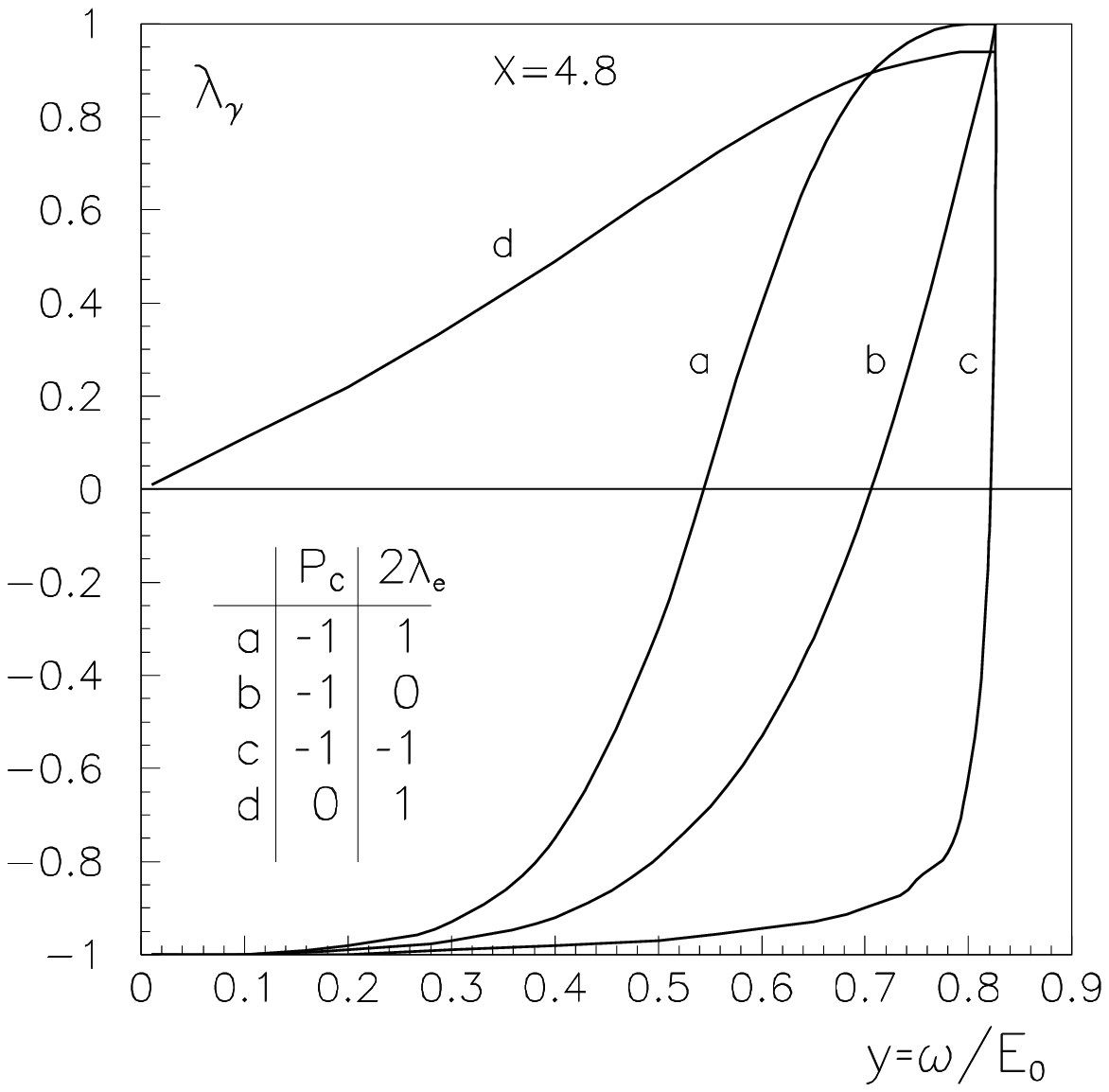, width=7.5cm}
\vspace*{-1.3cm}
\caption{ Helicity of scattered photons vs $\omega/E_0$ for various
polarizations of laser and electron beams.}
\label{fig6}
\end{minipage}%
\hspace*{0.6cm}\begin{minipage}[b]{0.45\linewidth}
\centering
\vspace*{-0.2cm} 
\hspace*{-0.75cm} \epsfig{file=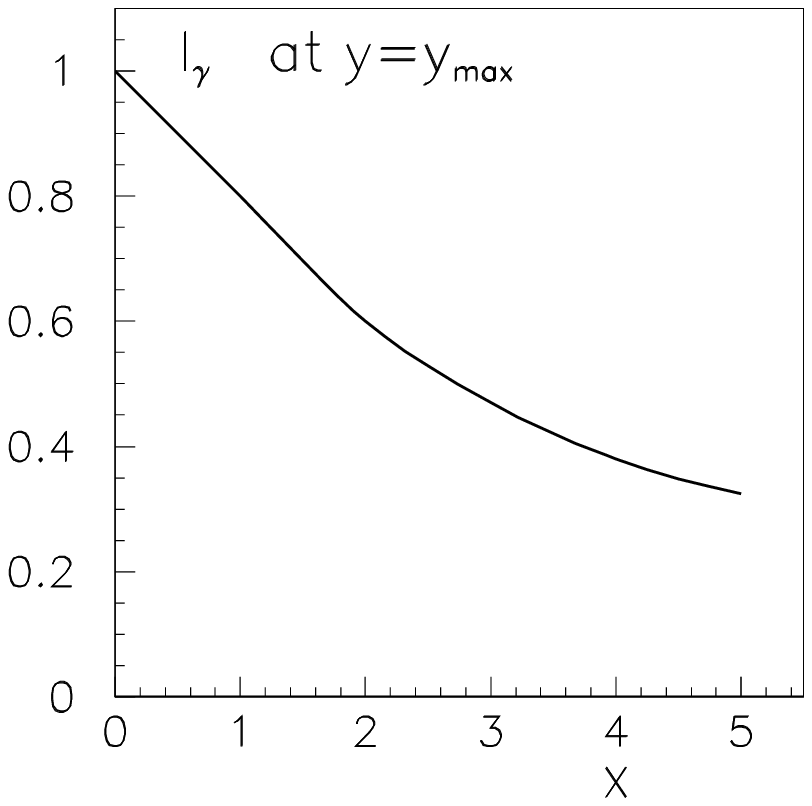,width=8.5cm,angle=0} 
\vspace*{-1.95cm}
\caption{Linear polarization of the scattered photons at $y=y_m$
vs x for $P_t=1$ and $\lambda_e=0$}
\label{linpol}
\end{minipage}
\end{figure}

  Using the polarized initial electrons and (or) laser photons, one
can obtain high energy photons with various polarizations~\cite{GKST84}.
In particular, the average helicity is
\begin{equation}
\lambda_\gamma(y)=\frac{2\lambda_e x r[1+(1-y)(1-2r)^2]+
P_c(1-2r)((1-y)^{-1}+1-y)}
{(1-y)^{-1}+1-y-4r(1-r)+2P_c\lambda_e rx(1-2r)(2-y)}.
\end{equation}
It is shown in fig.\ref{fig6} for $x=4.8$.  Note, if polarization of
laser photons $P_c = \pm 1,$ then $\lambda_\gamma = -P_c$ at $ y=y_m$.
In the case of $ 2P_c\lambda_e = -1$ (the case with good
monochromaticity) all the photons in the high energy peak have a high
degree a like-sign polarization. 
Note that for low $x$ the photon helicity
$\lambda_{\gamma}$ depends on $\lambda_e$ very slightly and high
energy photons have very high circular polarization in a wide energy
range near the maximum energy even at $\lambda_e = 0$.

  High degree of the circular photon polarization is essential for
suppression of the QED backgrounds for the Higgs production, because
$\sigma(\GG\ \to H) \propto 1+\lambda_{\gamma 1}\lambda_{\gamma
2}$, while the main  QED background $\sigma(\GG\ \to q\bar{q}) \propto
1-\lambda_{\gamma 1}\lambda_{\gamma 2}$.

The degree of the linear polarization of the Compton scattered photons is
\begin{equation}
l_{\gamma}=\frac{2r^2P_t}{(1-y)^{-1}+1-y-4r(1-r)+2P_c\lambda_e rx(1-2r)(2-y)},
\label{lg}
\end{equation}
where the linear polarization of laser photons $P_t$ and their
helicity $P_c$ for the case of complete polarization are connected by
relation $P_t^2+P_c^2=1$.

 The dependence of $l_{\gamma}$ at $y=y_m$ on the parameter $x$ is
shown in fig.\ref{linpol} for $P_t=1$ and unpolarized electron
beams. At large $x$ is not sufficiently high. It is of interest that
one can get larger $l_{\gamma}$, up to $l_{\gamma}=1$, it is so for
$2\lambda_eP_c=x(x+2)/(x^2+2x+2)$. However, in this case
$2\lambda_eP_c \approx +1$, that corresponds to curve $c$ in
fig.\ref{fig4}, when the number of photons with the energy near
$\omega_m$ is small.

Linear polarization of photon beams will be very useful for
determinating  the Higgs CP parity. The cross section of the Higgs
production by two photons $\sigma(\GG\ \to H) \propto 1 \pm
\l_{\gamma_1}\l_{\gamma_2}$ for CP$=\pm$ respectively. 

 General formulae for the polarization of scattered photons summed
over the polarization of the final electrons are given in
refs.\cite{McM,GKST84}. However, simulation the multiple
Compton scattering requires knowledge of final electron
polarization. This problem was considered recently in ref.\cite{Kot}.

Finally, it turns out that the polarization of electrons and high
energy photons traveling in a laser target is changed even for
particles which have not scattered. This unusual and important
phenomena will be discussed later.
\subsubsection{\EPEM\ pair production} \label{pairs}
\begin{figure}[!htb]
\vspace{-0.9cm}
\hspace*{0cm}\begin{minipage}[b]{0.45\linewidth}
\centering
\vspace*{-0.cm} 
\hspace*{0.cm} \epsfig{file=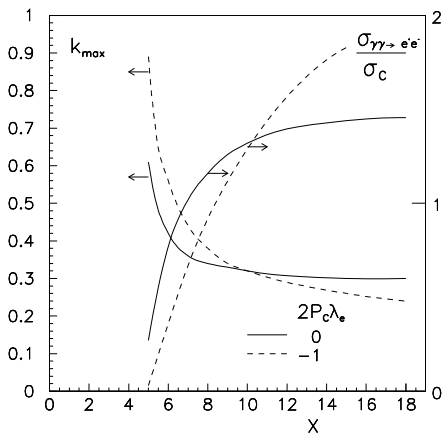,width=7.5cm,angle=0} 
\vspace{-0.4cm}\
%\label{linpol}
\end{minipage}%
\hspace*{1.1cm} \begin{minipage}[b]{0.4\linewidth}
\centering
\caption{a) The ratio of cross sections for \EPEM\ pair creation in
the collision of laser and high energy photons and for Compton
scattering and b) the maximum conversion coefficient vs $x$ 
assuming $\omega = \omega_m$.}
\vspace{2.5cm}     
\label{fig11}
\end{minipage}
\vspace{-1.2cm}
\end{figure}

 With increase of the laser photon energy the energies of the
scattered photons also increases and monochromaticity of the spectrum
improves.  However, besides Compton scattering, other processes become
possible in the conversion region~\cite{GKST83,TEL90}.  The
most important one is the process of \EPEM\ pair production in a
collision of a laser photon with  high energy scattered photon:
$\gamma_0 +\gamma \to \EPEM.$  The threshold of this reaction is
$\omega_m\omega_0 > m^2 c^4$, i.e. $x = 2(1+\sqrt{2}) \approx 4.83$,
the corresponding wavelength and laser photon energy are
\begin{equation}
 \lambda= 4.2 E_0 [\TEV]\;\; \mu m;\mbox{\hspace{1. cm}}
 \omega_0 = 0.3/E_0 [\TEV]\;\;\EV. 
\end{equation}
Above the threshold region  the pair production cross
section exceeds the Compton one by a factor of 1.5--2~\cite{TEL90,TEL95}, see
fig.\ref{fig11}.  Due to this fact the
maximum conversion coefficient at $x\sim$ 10--20  is limited by
35--25\% respectively. For these reasons it is preferable to work at
$x \approx 4.8$. Although, this limit is not absolute. Usually,
at $x<4.8$ we assume $k=1$ that corresponds to the conversion probability 
$1-e^{-1}=0.63$. The ratio of \GG\ luminosities
$L(x=10)/L(x=4.8)=(0.35/0.63)^2=0.3$. However, the luminosity spectrum
at $x=10$ is narrower~\cite{TEL95} by a factor 1.7 that is also important.

  Besides, \EPEM\ pairs can be produced in a collision of an
electron with a laser photon (Bethe-Heitler process): $e + \gamma_0
\to e + \EPEM\ $. However, at $x < 20$ its cross section is less than that of
the Compton scattering by two orders of magnitude~\cite{GKST83}.
\subsubsection{Conversion coefficient}
The conversion coefficient depends on the  energy  of  the
laser flash $A$ as
\vspace{-0.4cm}

\begin{equation}
     k  = N_\gamma /N_e \sim 1-\exp (-A/A_0 )\;\;\;(\sim A/A_0\;\;at\;\;
 A < A_0  ).
\label{k}
\end{equation}
\vspace{-0.5cm}

Let us estimate $A_0$. At the conversion region the r.m.s radius of
the laser beam depends on the distance $z$ to the focus (along the
beam) in the following way~\cite{GKST83}: $ r_\gamma = a_\gamma \sqrt{
1 + z^2 /Z_R^2}$, where the Rayleigh length $Z_R =2\pi a^2_\gamma
/\lambda , \; a_\gamma$ is the r.m.s. focal spot radius.  
The density of laser photons $n_\gamma =
(A/\pi r^2_\gamma \omega_0)\times$ $\exp (-r^2/r^2_\gamma) F_\gamma(z+ct),$
where $ \int F_\gamma(z) dz = 1$.

Let us assume the linear density of photons to be uniform along  the
beam:  $F_\gamma = 1/l_\gamma$ and $Z_R \ll l_\gamma$. The
probability of the Compton scattering  for the electron moving through a
laser target along the axis~\cite{TEL95}
\begin{equation}
p=2 \int n_\gamma \sigma_c dz \sim \frac{2A\sigma_c}
{\pi a^{2}_{\gamma}\omega_0 l_\gamma} \int\limits_{-\infty}^{\infty}\frac{dz}
{1+\frac{z^2}{Z_R^2}} = \frac{2A\sigma_c}{c \hbar l_\gamma}.
\end{equation}
 One collision length corresponds to $ p = 1$
that gives   
%\begin{equation}
$A_0 = \hbar c l_\gamma/2\sigma_c,  
P=A_0c/l_\gamma = \hbar c^2/2\sigma_c.$
%\end{equation}

 The minimum laser flash energy (corresponding to $l_\gamma = l_e$)
and power needed for obtaining the conversion coefficient $k \sim
63\% \;$ ($A=A_0$) at $x =4.8$ $(\sigma_c \approx
1.9\times10^{-25}\,\CM^2)$
\begin{equation}
A_{0,min} = \hbar c l_e/2\sigma_c = 8.4 \;l_e[\CM] \;\;J, \;\;\;\;
P_{0,min} =  \hbar c^2/2\sigma_c \approx 2.5 \times 10^{11} \;\;W.
\end{equation}
These minimum values for $ A_0$ and $ P_0$ have been obtained for
uniform photon density distribution along the electron and laser beams and
$Z_R \ll l_e =l_\gamma$.The value of $A_0$ is almost independent of the
focal spot size at $ 2Z_R < l_e =l_\gamma$, i.e.
$a_\gamma < \sqrt{\lambda l_e/4\pi}$.  For Gaussian beams with $l_e =l_\gamma
\equiv 2\sigma_z \gg Z_R$  calculations gives~\cite{TEL95} 
$A_0\approx \sqrt{\pi} A_{0,min}$  and $P_{peak}\approx \sqrt{2} P_{0,min}$

   Let us consider another example:\, $2Z_R \approx l_\gamma \ge
l_e$ when $A_0$ is only slightly larger than $A_{0,min}$,
but the laser target has the relatively low density that is important
for avoiding multiphoton processes.  In this case, the radius of laser beam
$ r_\gamma \sim a_\gamma$ along the target and the density of laser
photons $ n_\gamma \sim A/(\pi \omega_0 a_{\gamma}^{2} l_\gamma )$.
The probability of conversion $p \sim n_\gamma \sigma_c l_\gamma=1$ at
\begin{equation}
A_0 \sim \pi\hbar cl_\gamma/2\sigma_c = \pi A_{0,min} (l_{\gamma}/l_e).
\label{a0g}
\end{equation}
\subsubsection{Nonlinear effects in Compton scattering
      in the conversion region.}\label{nonlinear}
\IND\ In the conversion region the density of laser photons can be
very high that leads to multiphoton processes~\cite{LANDAU}-\cite{BaiYok}
which were studied recently at SLAC:~\cite{BULA}
$$ \mbox{e}+n\gamma_0 \to \mbox{e} +\gamma,\,\,\,\,\gamma +n\gamma_0
 \to \EPEM, $$ where $e,\gamma$ are high energy electron and photon
 respectively, $ \gamma_0$ denotes a laser photon.  Nonlinear effects
depend  (beside  $x$) on the parameter
\begin{equation}
\xi^2 = (eB\hbar/m\omega_0 c)^2,
\end{equation}
where $B$ is the field strength, $\omega_0$ is the photon
energy.  At $\xi^2 \ll 1$ an electron interacts with one photon 
(Compton scattering), while at $\xi^2 \gg 1$ an electron
interacts with a  collective  field (synchrotron radiation).

   What values of $\xi^2$ are acceptable? In a strong field, an electron
has transverse motion, which increases the effective electron 
mass:~\cite{LANDAU} $m^2 \to m^2 (1+\xi^2)$. Due to this fact the
maximum energy of scattered photons is decreased by $\Delta
\omega_m/\omega_m = \xi^2/(x+1)$, that is 5\% at $\xi^2 =0.3$ and $x=4.8$.
Let us estimate  $\xi^2$ in  our case. Assuming $2Z_R \approx
l_\gamma$ (as for eq.\ref{a0g}) 
and taking into account that the field in the
laser focus is $B^2 =4A/a_{\gamma}^{2} l_\gamma$ and $A=kA_0$
one get~\cite{TEL90,TEL95}
\begin{equation} \xi^2 \approx
\frac{2}{\pi\alpha} \frac{\sigma_0}{\sigma_c}
 \frac{\lambda}{l_\gamma} k
\end{equation}
At $x=4.8$  and  $k=1$ we get $\xi^2 = 0.05 E_0 [\TEV]/l_\gamma.$
For example, if $l_\gamma = l_e = 200\, \MKM\,\,\,$ and $E_0
=1\,\, \TEV,$  then $ \xi^2 =2.5$.  This is not
acceptable.  To decrease $\xi$, keeping $ k=const$, one has to increase 
 $ l_\gamma$, correspondingly the flash energy should be 
proportionally higher (eq.\ref{a0g}).

   In principle, there is solution of this problem, it is {\it
stretching}  the laser focus.~\cite{TSB1}. One can focus $n$ separate
lasers to $n$ adjacent conversion regions along the electron beam
pathway. The total flash energy needed for the conversion remains the
same, but $\xi^2$ is smaller than that with one conversion region by a
factor of $n$. Another way, stretching a ``chirped'' pulse, is
discussed in sect.\ref{lascool}.
\subsubsection{Variation of high energy electrons and photons polarization
in laser wave.}\label{variation}
{\it Polarization of unscattered electrons}
\vspace{3mm}

   Let us consider first the following example: an unpolarized
electron beam is collided with a circularly polarized laser
pulse. Some electrons pass this target without the Compton scattering.
What happened with their polarization? It is changed, since the cross
section of the Compton scattering depends on the product
$P_c\lambda_e$ and unscattered electron beam contains already unequal
amount of electrons with the forward and backward helicities. Considering
the multiple Compton scattering  we should take this effect into
account.
    
General formulae for this effect has been obtained in
ref.\cite{KSelectron} (see V.Serbo talk at this Workshop), where the
variation in polarization of the unscattered electrons was considered
as the result of interference of the incoming electron wave and the
wave scattered at the zero angle.

\vspace{3mm}
\noindent{\it Variation of $\gamma$-quanta polarization in a laser wave.} \vspace{3mm}  

  It is well known that a region with an electromagnetic field can be
regarded as an anisotropic medium~\cite{LANDAU}. Strong laser field
also has such properties. As a result, the polarization of  high
energy photon produced in the Compton scattering may be changed during
its propagation through the polarized laser target~\cite{KSphoton}
(see V.Serbo talk at this Workshop). This effect is large at $x
\approx 4.8$ (the threshold for \EPEM\ production).  Note, that in the
case most important for us  $2P_C\lambda_e=1$ the polarization of high
energy circularly polarized photons is not changed when they propagate
in the circularly polarized laser wave.

In principle, using two adjacent conversion regions one can produce
first circular polarized photons (using a circularly polarized laser)
and then change the circular polarization to the linear one using a linearly
polarized laser.

  A similar effect, influence of the electromagnetic field on the
polarization of high energy photons, exists also at the interaction
region of photon colliders~\cite{TELvav}. 
\section{Effects at the interaction region, ultimate \GG\ luminosity}
 Due to the absence of beamstrahlung, beams in \GG\ collisions can
have much smaller horizontal beam sizes than  in \EPEM\ collisions.
 However, even after optimization of the final focusing
system the attainable \GG\ luminosity in current LC projects is
determined by the attainable ``geometric'' ee--luminosity.
 For the ``nominal'' beam parameters (the same as in
 \EPEM\ collisions) and  optimum final focus systems the
   luminosity $\LGG(z>0.65) \sim 10^{33}$ \CMS\ ($z=W_{\GG}/2E_0$.
 Obviously, this is not a fundamental limit.

 The only collision effect restricting \GG\ luminosity at photon
  colliders is the coherent pair production which leads to the
  conversion of a high energy photon into \EPEM\ pair in the field of
  the opposing electron beam~\cite{CHEN,TEL90,TEL95}. The probability
  of this process is high when $B \geq B_{cr}=\alpha e/\gamma r^2_e=
  2.2\times10^{7}/E$[\TEV] G.

 There are three ways to avoid this effect: a) use flat beams; b)
 deflect the electron beam after conversion at a sufficiently large
 distance from the IP; c) under certain conditions (low beam energy,
 long bunches) the beam field at the IP is below the critical one due
 to the repulsion of electron beams~\cite{TELSH}. The problem of
 ultimate luminosities for different beam parameters and energies was
 analyzed recently in ref.\cite{TSB2} analytically and by simulation.
 The resume is following.  At ``low'' energies ($2E< 0.3-1\; \TEV$,
 depending on the beam length), the coherent pair production is not very
 important, even for a very small $\sigma_x$'s, it is because electron
 beams are repulsing each other so that the field on the beam axis
 (which affects high energy photons) is below the critical field. It
 means that the $\gamma\gamma$ luminosity is simply proportional to
 the geometric electron-electron luminosity $\LGG(z>0.65)\sim 0.1
 L_{ee}$.  Having electron beams with sufficiently low  emittances one can
 obtain, in principle, $\LGG(z>0.65)$ above $10^{35}$ \CMS.  

  One of the main problems here is the generation of electron beams with
small emittances in both horizontal and vertical directions.  One of
promising techniques is discussed below.

\section{Laser cooling of electron beams}\label{lascool}
The transverse beam sizes at the interaction point are determined by
the emittances \EX, and \EY: $\sigma_i=\sqrt{\EI\ \BI\ }$, 
where \BI\ is the beta function
at the IP.  With an increase of the beam energy the emittance of the
bunch decreases: $\EI=\ENI/\gamma$, where $\gamma=E/mc^2, $ \ENI\ is
the {\it normalized} emittance.  The beams with a small \ENI\ are
usually prepared in damping rings which naturally produce bunches with
$\ENY\ll\ENX$.  Laser RF photoguns can also produce beams with low
emittances.  However, for linear colliders it is desirable to have
smaller emittances.  Recently a new method of electron beam cooling
was proposed -- laser cooling -- which allows further reduction of the
transverse emittances after damping rings or guns by 1--3 orders of
magnitude~\cite{TSB1}.
\subsection{One pass laser cooling}
The idea of laser cooling of electron beams is very simple.  During a
collision with optical laser photons (in the case of strong fields it
is more appropriate to consider the interaction of an electron with an
electromagnetic wave) the transverse distribution of electrons
($\sigma_i$) remains almost the same. Also, the angular spread
($\sigma_i^{\prime}$) is almost constant, because electrons loss  momenta
 almost along their
trajectories (photons follow the initial electron trajectory with a
small additional spread). So, the emittance $\EI = \SI \SIP$ remains
almost unchanged. At the same time, the electron energy decreases from
$E_0$ down to $E$. This means that the transverse normalized
emittances have decreased: $ \EN = \gamma \E = \EN_0(E/E_0)$.  One can
reaccelerate the electrons up to the initial energy and repeat the
procedure. Then after N stages of cooling $ \EN /\EN _0 = (E/E_0)^N$
(if \EN\ is far from its limit).
There are several questions to this method: 1)
requirements on laser parameters 
2) an energy spread of the beam after cooling 3) a limit on
the final normalized emittances;
4) depolarization.
\subsubsection{Flash energy.}
Physics in the cooling region is the same as in the conversion region
of photon colliders sect.\ref{conv}. Typical values of parameters in
this problem: $E_0 \sim 5\;\GEV, x \ll 1,\; \xi^2 \sim 0.5 - 10$.

Passing the laser target the electron gradually loss 
its energy and at the exit~\cite{TSB1}
(it is assumed that $Z_R \ll l_{\gamma} \simeq l_e$)
\begin{equation}
 \frac{\EN_0}{\EN} \simeq \frac{E_0}{E} = 
1+\frac{64\pi^2r_e^2\gamma_0}{3mc^2\lambda\l_e}A; \;\;\;
 A[J] = \frac{25\lambda[\MKM\ ]l_e[\MM\ ]}{E_0[\GEV\ ]}
\left(\frac{E_0}{E}-1\right).
\label{c3}
\end{equation}
 For example: at $\lambda=0.5\; \MKM,\; l_e=0.2\; \MM,
E_0=5\;\GEV, E_0/E=10$ the required laser flash energy $A = 4.5\;$ J.
To reduce the laser flash energy in the case of long electron bunches,
one can compress the bunch  before cooling as much as possible
and stretch it after cooling up to the required value.

 Assume $Z_R \sim 0.25l_e$ we can find the parameter $\xi^2$: 
\begin{equation}
\xi^2 = \frac{16r_e\lambda A}{\pi l_e^2 mc^2} =
\frac{3\lambda^2}{4\pi^3 r_e l_e \gamma_0}\left(\frac{E_0}{E}-1\right
) \ = 4.3 \frac{\lambda^2 [\MKM\ ]}{l_e [\MM\ ]E_0 [\GEV\ ]}
\left(\frac{E_0}{E}-1\right ).
\label{c5}
\end{equation}  
In the previous example $\xi^2$ = 9.7. 
In principle, both the ''undulator'' and ''wiggler'' cases are possible.
We will see later that in order to have lower limit on emittance and
smaller depolarization it is necessary to have a low $\xi^2$. With a
conventional  optics one can reduce $\xi^2$ only by increasing $l_{\gamma}$
(and $Z_R$) with a simultaneous increase of the laser flash
energy. From (\ref{c3}) and (\ref{c5}) we get
$A \propto (\lambda^3/\gamma_0^2 \xi^2)(E_0/E-1)^2$.

In sect.\ref{conv} we already discussed the way of reducing $\xi^2$
keeping the flash energy constant, it is stretching of the conversion
region without changing the radius of this area.  One of possible
solutions~\cite{TSB1} uses chirped laser
pulses (wave length depends linearly  on longitudinal
position) and chromaticity of the focusing system.  In this scheme,
the laser target consists of many laser focal points (continuously)
and light comes to each point exactly at the moment when the electron
bunch is there. The required flash energy is determined only by
diffraction and at the optimum wave length it does not depend on the
collider energy.

The electron energy spread   arises from the 
quantum-statistical nature of radiation.  After energy loss 
$\Delta E$, the increase of the energy spread
$\Delta(\sigma_E^2)=\int\varepsilon^2 \dot{n}(\omega) d\omega dt
=-aE^2\Delta E$,  where $\dot{n}(\omega)$ is the spectral density
of photons emitted per unit time, $a=14\omega_0/5m^2c^4=7x_0/10E_0$ for
the Compton case and $a=55\hbar e B_0/(8\pi\sqrt{3}m^3c^5)\;$ for the
``wiggler'' case.
   
There is the  second effect which leads to  decreasing  the energy
spread.  It is due to the fact that $dE/dx \propto E^2$ and an electron
with higher (lower) energy than the average  loses more (less) than on
average.  This results in the damping: $d(\sigma_E^2)/\sigma_E^2 =
4dE/E $ (here $dE$ has negative sign). The full
equation for the energy spread is $d\sigma_E^2 = -aE^2dE +
4(dE/E)\sigma_E^2$, with  solution
$\sigma_E^2/E^2 = \sigma_{E_0}^2E^2/E_0^4+
aE_0(E/E_0)(1-E/E_0)$. In our case
\begin{equation}
\frac{\sigma_E^2}{E^2}  \sim\frac{\sigma_{E_0}^2E^2}{E_0^4}+
  \frac{7}{10}x_0(1+\frac{275\sqrt{3}}{336\pi}\xi)\frac{E}{E_0}
  \left(1-\frac{E}{E_0}\right),
\label{c7}
\end{equation}
here the results for the Compton scattering and SR are joined together.
Example: at $\lambda=0.5\;\MKM,\; E_0=5\; \GEV\ (x_0 = 0.19)$ and
$E_0/E=10$,  the  Compton term alone gives $\sigma_E/E \sim 0.11$
and with the ``wiggler'' term ($\xi^2 = 9.7$, see the example above)
$\sigma_E/E \sim 0.17$.
What $\sigma_E/E$ is acceptable? In the last example $\sigma_E/E \sim
0.17$ at E = 0.5 GeV (after  cooling). This means that at the
collider energy E = 250 GeV we will have $\sigma_E/E \sim 0.034\%$,
that is better than necessary (about 0.1 \%).

In a two stage cooling system, after reacceleration to the initial
energy $E_0$ = 5 GeV the energy spread is $\sigma_E/E_0 \sim
1.7\%$. For this value there may be a problem with focusing of
electrons which can be solved using a focusing scheme with 
correction of chromatic aberrations.  
\subsubsection{The minimum normalized emittance} 
It is determined by the quantum nature
of the radiation. Let us start with the case of  pure Compton
scattering at $\xi^2 \ll 1$ and $x_0 \ll 1$. In this case, the
scattered photons have the energy distribution (see eq.\ref{difspec}:) $dp =
(3/2)[1-2\omega/\omega_m +2(\omega/\omega_m)^2]d\omega/\omega_m$, 
where $\omega_m = 4\omega_0\gamma^2$. The angle of
the electron after scattering is \cite{GKST83} $\theta_1^2 =
(\omega_m\omega - \omega^2)/(\gamma^2E^2)$. After averaging over the
energy spectrum we get the average $\theta_1^2$ in one collision:
$\langle\theta_1^2\rangle = 12\omega_0^2/(5m^2c^4)$. After many Compton
collisions ($N_{coll}$) the r.m.s. angular spread in i=x,y projection
$\Delta\langle\theta^2_i\rangle = 0.5\Delta\langle\theta^2\rangle =
0.5N_{coll}\langle\theta_1^2\rangle = -0.5(\Delta
E/\bar{\omega})\langle\theta_1^2\rangle = -3\omega_0\Delta E/5E^2$.

The normalized emittance $\ENI{^2} = (E^2/m^2c^4)\langle r_i^2\rangle
\langle\theta_i^2\rangle$ does not change when $\Delta
\langle\theta_i^2\rangle/ \langle\theta_i^2\rangle$ = $-2\Delta E/E.$
Taking into account that
$\langle\theta_i^2\rangle\equiv\ENI\ /\gamma\beta_i$ we get the
equilibrium emittance due to the Compton scattering
%]
\begin{equation}
\ENI_{, min} \approx
 0.5\gamma E \beta_i\Delta \langle\theta_i^2\rangle/\Delta E =
 \frac{3\pi}{5}\frac{\lambda_c}{\lambda}\beta_i 
 =\frac{7.2\times 10^{-10}\beta_i [mm]}{\lambda [\MKM\ ]} \;\;
  \mbox{m$\cdot$rad},
\label{c8}
\end{equation}
where $\lambda_c=\hbar/mc$.  For example: $\lambda =0.5\; \MKM,
\;\beta=l_e/2=0.1\;\MM\ ($NLC$) \Rightarrow
\EN_{,min}=1.4\times10^{-10}\;\M\cdot$rad. For comparison in the NLC
project the damping rings have $\ENX\ =3\times 10^{-6}\; \M\cdot$rad,
$\ENY\ =3\times 10^{-8}\; \M\cdot$rad.

If $\xi^2 \gg 1$,  the electron moves as
in a  wiggler. Assuming that the wiggler is planar and deflects the
electron in the horizontal plane  one can obtain 
the equilibrium normalized emittance~\cite{TSB1}
\begin{equation}
\ENX = \frac{11e^3\hbar c \lambda^2 B_0^3
\beta_x}{24\sqrt{3}\pi^3(mc^2)^4} =
\frac{11}{3\sqrt{3}}\frac{\lambda_C}{\lambda}\beta_x \xi^3 \approx
 \frac{8\times 10^{-10}\beta_x [\MM\ ] \xi^3}{\lambda [\MKM\ ]} \;\;
\mbox{m$\cdot$rad}.
\label{c9}
\end{equation}
  For our previous example we have $\xi^2 =9.7$ and $\ENX\ =
5\times 10^{-9}\;\M\cdot$rad (in the NLC \ENX\ = 3$\times 10^{-6}$
m$\cdot$rad). Stretching  the cooling region with \n=10, further
decreases the horizontal emittance by a factor 3.2.

 For the minimum vertical normalized emittance at
$\xi^2 \gg\ 1$ estimations give the value~\cite{TSB1}
\begin{equation}
\ENY_{min} \sim 3\left(\frac{\lambda_c}{\lambda}\right)\beta_y\xi 
\approx \frac{1.2\times 10^{-9}\beta_y [\MM\ ] \xi}{\lambda [\MKM\ ]} \;\;
\mbox{m$\cdot$rad}.
\label{c10}
\end{equation}   
For arbitrary $\xi^2$ the minimum emittances can be estimated as the sum
of (\ref{c8}) and (\ref{c9}) for \ENX\ and sum of (\ref{c8}) and (\ref{c10})
for \ENY\
\begin{equation}
  \ENX \approx
  \frac{3\pi}{5}\frac{\lambda_C}{\lambda}\beta_x(1+1.1\xi^3);\;\; \ENY
  \sim \frac{3\pi}{5}\frac{\lambda_C}{\lambda}\beta_y(1+1.6\xi).
\label{c11}
\end{equation}

Finally, the depolarization of the electron beam during the laser
cooling~\cite{TSB1}
\begin{equation} 
 \Delta\zeta/\zeta \approx 0.3x_0(1+1.8\xi). 
\label{c14}
\end{equation}
For the previous example with $\xi^2 =9.7$ and $x_0 = 0.19$ we get
$\Delta\zeta/\zeta = 0.057+0.32 = 0.38$, that is not acceptable. This
example shows that the depolarization effect imposes the most
demanding requirements on the parameters of the cooling system. The
main contribution to depolarization gives the second term. 
Stretching the cooling region by a factor of ten
we  can get $\Delta\zeta/\zeta = 0.057+0.1\sim 15\%$.

Possible sets of parameters for the laser cooling is the following:
$E_0 = 4.5$ GeV, $l_e=0.2 $ mm, $\lambda = 0.5$ \MKM, flash energy $A
\sim 5-10 $ J, focusing system with stretching factor \n=10. The final
electron bunch will have an energy of 0.45 \GEV\ with an energy spread
$\sigma_E/E \sim 13 \%$, the normalized emittances \ENX,\ENY\ are
reduced by a factor 10, the limit on the final emittance is $\ENX\
\sim\ENY\ \sim2\times 10^{-9}\;$ m$\cdot$rad at $\beta_i= 1\; \MM$,
depolarization $\Delta\zeta/\zeta \sim 15\%$.  
  The two stage system with the same parameters gives
100 times reduction of emittances (with the same limits).

For the cooling of the electron bunch train one laser pulse can be
used many times.  According to (\ref{c3}) $\Delta E/E = \Delta A/A$ and even
25\% attenuation of laser power leads only to small additional energy
spread.

The proposed scheme of laser cooling of electron beams seems very
promising for future linear colliders (especially for photon
colliders) and allows to reach ultimate luminosities.  Perhaps this
method can be used for X-ray FEL based on high energy linear
colliders.
\subsection{Laser cooling in storage rings}
  Application of the laser cooling for damping rings has
one problem: an electron loses in each Compton scattering a
large amount of its energy (up to $4\gamma^2\omega_0$) that leads to
very large beam energy spread or knock out of the electrons from the
beam. Nevertheless,  Huang and Ruth have shown~\cite{Huang} that
the laser cooling can   successfully be used  for  storage rings
in the energy range from a few MeV to a few hundred MeV.

The minimum transverse emittance in a laser cooled storage ring is
given by eq.\ref{c8}: For example, for $\lambda=1\;$\MKM\ and $\beta =$
1 cm  the minimum normalize emittance is about $7.3\times10^{-9}$
m rad, much better than that given  by  other sources.

 The energy spread is determined by fluctuations in the Compton
scattering.  For the Compton scattering (see two paragraphs above
eq.\ref{c7}) $\Delta(\sigma_E^2)=-(14\omega_0/5m^2c)E^2dE$. In
equilibrium this growth of emittance is compensated by the radiation
damping $d(\sigma_E^2)/\sigma_E^2 = 2dE/E $ (due to synchrotron
oscillations in storage rings the cooling rate is smaller than that in
linear cooling by a factor of 2. By comparing heating and cooling rate
one gets~\cite{Huang}
\begin{equation}
  \sigma_E/E = \sqrt{14\pi\lambda_c\gamma/5\lambda}, 
\;\;\;\;\lambda_C=\hbar/mc.
\label{r2}
\end{equation}  
  For example, when $E=100$ MeV and $\lambda=$ 1 \MKM\ the r.m.s
energy spread is about 2.6 \%. The momentum acceptance for such ring
should be about 15\%.

   Beside effects in the Compton scattering, we should not forget the
beam space charge and intrabeam scattering problems inherent to low
energy rings. In ref.\cite{Huang} two configuration were considered:
a) E $=100 \;$MeV$, \;$N $=1.3\times10^{10},\; $ R $= 1$ m; b) E $=8 \;
$MeV$,\; $N $=1.1\times10^{10},\; $R $= 0.5$ m. 
For these parameters they obtained the
equilibrium normalized emittances of $1\times10^{-7}$ m and
$2\times10^{-5}$ m  or by factor 15 and
3000 larger than it is following from eq.\ref{c8}. For more details
see ref.\cite{Huang} and Huang and Ruth talks at this workshop.
   In any case, laser cooling is a very new field and requires more
detailed study. Both schemes of cooling are attractive and have 
complimentary areas of application.  
\section{Positron production using a laser}
Production of positrons for linear \EPEM\ colliders is a challenging
problem. Several possible solutions are based on a laser technique:

1. Positron production by high energy electron beam in a laser wave. 

2. Two step scheme, similar to that of ref.\cite{BalMikch}, but instead of
an undulator a laser wave is used. 
\subsection{Positron production by electrons in a laser wave}
   In the section \ref{conv} we have seen that high energy electrons
at the conversion region can produced \EPEM\
pairs~\cite{GKST83,TEL90}. This is the a two-step process: 1) the high
energy photon is produced in the Compton scattering of a laser photon
on  high energy electron, 2) \EPEM\ pair is produced in collisions of
high energy photon with a laser photon. The threshold of this
process is $x=4.8$.  Baier~\cite{Baier97} has considered this scheme
for the production of polarized positrons. Practical use of this scheme is
a big question. It requires electron beams with high energy ($E_e
\sim 60$ GeV for laser with $\lambda=0.25$ \MKM), the produced
positrons have  large energies and {\it too large energy spread}.

  Chen and Palmer~\cite{ChenPalmer} proposed to use the same scheme
 but with a very strong laser field ($\eta>1$ and $\Upsilon >1$). In
 this case the $\gamma,e^+,e^-$ shower is developed which stops when
 the photon energies of  became below the coherent pair production
 threshold. As a result, one can obtain (one of examples) $N_{e^+}\sim
 3N_{e}$ with the energy about 1.5 GeV and 7\% r.m.s energy
 spread. It is remarkable that the normalized emittance of the final
 beam can be much lower than that of the initial electron beam.
 Unfortunately, final positrons and electrons in this scheme are
 unpolarized. This method also requires high energy initial electron
 beams (above 50 GeV).
\subsection{Two step scheme: Compton scattering + W-target.}
  This scheme was described recently by Hirose, Omori, Tsunemi et
al. at LC97 in Zvenigorod and it is prepared now for testing at KEK. The
6.7 GeV electron beam is collided with a focused circularly polarized
CO$_2$ laser  and produces photons with the maximum energy of 80
MeV. Photons in the high energy part of spectrum have the high degree of the
circular polarization. Further these photons are brought to the W
target where \EPEM\ pairs are produced.  Positrons with $E_{e^+}
> 35$ MeV have the average longitudinal polarization about 80 \%.
What is essential in this scheme:

1. Each electron produces  more than one (about 6)  photons.

2. The laser target consists of 40 conversion regions (created by 40
lasers) with 5 cm spacing. This is done  for two reason: a) it is
a way to create a thick laser target using several low power lasers, b) in
this case, the nonlinear parameter $\xi^2 \ll 1$ (an electron is
scattered on one laser photon). This is very important to have
collisions with only one laser photon. If the number of absorbed laser
photons is fluctuating the maximum energy of scattered photons is
fluctuating as well, polarization is no longer an unique function of
the photon energy and as a result, the photons at any energy have low degree
of polarization.

The total flash energy of 40 CO$_2$ lasers in the KEK scheme is about
10 J, each electron produces 5.7 photons with the energy 1-80 MeV, the
yield of polarized positrons per one initial electron is about 0.2.
For the JLC positron injector the total power consumption from plug is 17 MW 
for linac and 3.2 MW for the laser system. Numbers are reasonable.

\vspace{-0.5cm}  


\begin{thebibliography}{99}
%
\bibitem{LOEW} G.~Loew et al., {\it Int. Linear Collider Tech.
Rev. Com. Rep.}, SLAC-Rep-471(1996).
\vspace{-0.2cm}
%
\bibitem{GKST81} I.Ginzburg, G.Kotkin, V.Serbo, V.Telnov, {\it
Pizma ZhETF}, {\bf 34} (1981) 514; {\it JETP Lett.}{\bf 34} (1982) 491.
\vspace{-0.2cm}
%
\bibitem{GKST83} I.Ginzburg, G.Kotkin, V.Serbo, V.Telnov, {\it Nucl. Instr.
and Meth.} {\bf 205} (1983) 47.
\vspace{-0.2cm}
%
\bibitem{GKST84} I.Ginzburg, G.Kotkin, S.Panfil, V.Serbo, V.Telnov,
 {\it Nucl. Instr. and Meth.} {\bf 219} (1984) 5.
\vspace{-0.2cm}
%
\bibitem{TEL90} V.Telnov, {\it Nucl. Instr. and Meth.A} {\bf 294} (1990) 72.
\vspace{-0.2cm}
%
\bibitem{TEL95} V.Telnov, {\it Nucl. Instr. and Meth.A} {\bf 355} (1995) 3.
\vspace{-0.2cm}
%
\bibitem{BERK} {\it Proc.of Workshop on \GG\ Colliders},
Berkeley CA, USA, 1994, {\it Nucl. Instr. and Meth. A}
 {\bf 355} (1995) 1--194.
\vspace{-0.2cm}
%
\bibitem{TELee} V.Telnov, {\it Proc. of 2 nd Inter. Workshop on $e^-e^-$
inter. at TeV energies.}  Santa Cruz,  USA,
Sept.22-24, 1997; to be published in Intern. J. of Mod. Phys. A,
preprint Budker INP-98-02, e-print hep-ex/9802003.
\vspace{-0.2cm}
%
\bibitem{NLC} {\it Zeroth-Order Design Report for the Next Linear Collider} 
LBNL-PUB-5424, SLAC Report 474, May 1996.\vspace{-0.2cm}
\vspace{-0.2cm}
%
 
%
\bibitem{TESLA} {\it Conceptual Design of a 500 GeV Electron Positron
Linear Collider with Integrated X-Ray Laser Facility} DESY 79-048,
ECFA-97-182.\vspace{-0.2cm}

\bibitem{JLC} {\it JLC Design Study}, KEK-REPORT-97-1, April 1997.
\vspace{-0.2cm}
%
\bibitem{McM} W.H.McMaster, {\it ReV. Mod. Phys.} {\bf 33} (1961) 8;
 V.P.Gavrilov, I.A. Nagorskaya and V.A.Khose, {\it Izvestiya AN
Armenian SSR}, Fizika {\bf 4} (1969) 137; Ya.T.Grincishin, {\it
Yad.Fiz. (USSR)} {\bf 36} (1982) 1450.\vspace{-0.2cm}
%
\bibitem{Kot} G.Kotkin, S.Polityko, V.Serbo, {\it Yad. Fizika}
(russian), {\bf 59} (1996) 2229.
\vspace{-0.2cm}
\bibitem{LANDAU} V.Berestetskii, E.Lifshitz and L.Pitaevskii, {\it Quantum
Electrodynamic}, Pergamont press, Oxford, 1982.
\vspace{-0.2cm}
\bibitem{Nikishov}  A.I.Nikishov and V.I.Ritus, {\it Sov. Phys JETP,}
{\bf 19} (1964) 529.
\vspace{-0.2cm}
\bibitem{Narozhny} V.B. Narozhny, A.I.Nikishov and V.I.Ritus, {\it
Sov. Phys JETP,} {\bf 20} (1964) 622.
\vspace{-0.2cm}
\bibitem {RN} V.I.Ritus, A.I.Nikishov. Trudy FIAN, {\bf 111} (1979) [in
Russian].
\vspace{-0.2cm}
\bibitem{Baier81} V.N.Baier, V.M.Katkov, V.M.Strakhovenko, {\it
Sov. Phys JETP,} {\bf 53} (1981) 688.
\vspace{-0.2cm}
\bibitem{BaierCr} V.N.Baier, V.M.Katkov, V.M.Strakhovenko,
{\it Electromagnetic processes at high energy in oriented single
crystals}, World Scientific Co, Singapore.
\vspace{-0.2cm}
\bibitem {GKP} I.F.Ginzburg, G.L.Kotkin, S.I.Polityko, {\it Sov. Yad.
Fiz.} {\bf 37} (1983) 368; {\bf 40} (1984) 1495.
%
\vspace{-0.6cm}
\bibitem{Tsai} Yu.S.Tsai, {\it Phys. Rev.} {\bf D48} (1993) 96. 
\vspace{-0.2cm}
\bibitem {GrR} Ya.T. Grinchishin, M.P.Rekalo, {\it ZhETF} {\bf 84}
 (1983) 1605, Errata: {\it ZhETF} {\bf 86} (1984) 784; {\it
 Sov. Yad. Fiz.} {\bf 40} (1984) 181.
\vspace{-0.2cm}
\bibitem{Galynskii} M.V. Galynskii, S.M. Sikach, {\it
Zh.Eksp.Teor.Fiz.} {\bf 101} (1992) 828, 
{\it Sov. Phys. JETP} {\bf 74} (1992)
441-446.
\vspace{-0.2cm}
\bibitem{BaiYok} V.N.Baier, K.Yokoya, 
{\it Particle Accelerators}, {\bf 44} (1994) 77.
\vspace{-0.2cm}
\bibitem{BULA} K.T.Bula, E.J.McDonald et al.. {\it Phys. Rev. Lett.}
 {\bf 76} (1996) 3116.
\vspace{-0.2cm}
\bibitem{KSelectron} G.L.Kotkin, H.Perlt, V.G.Serbo, {\it
Nucl.Instr. and Meth.A} {\bf 404} (1998) 430.
\vspace{-0.2cm}
\bibitem{KSphoton} G.L.Kotkin, V.G.Serbo, {\it Phys. Lett.} {\bf B 413}
(1997) 122.
\vspace{-0.2cm}
\bibitem{TELvav} V.Telnov,  Proc. of Intern. Vavilov's Conference 
on nonlinear optics, Novosibirsk,
June 24-26, 1997; Budker INP 97-71,  physics/9710014.
\vspace{-0.2cm}
%
\bibitem{CHEN} P.Chen, V.Telnov, {\it Phys. Rev. Letters}, {\bf 63}
(1989) 1796.
\vspace{-0.2cm}
\bibitem{TELSH} V.Telnov,
{\it Proc.of Workshop``Photon 95'',} Sheffield, UK, April 1995, p.369.
\vspace{-0.2cm}
%
%
\bibitem{TSB2} V.Telnov,{\it Proc. of ITP Symp. on Future High Energy
Colliders,} Santa Barbara, USA, Oct. 21--25, 1996; AIP Conf. Proc. No
397, ed. Z.Parza, (AIP. New York 1997), p.259-273; Budker INP 97-47,
e-print: physics/9706003
\vspace{-0.2cm}
%
\bibitem{TSB1} V.Telnov, {\it Proc. of ITP Workshop ``New modes of
particle acceleration techniques and sources''} Santa Barbara, USA,
August 1996, AIP conf. proc. 396, SLAC-PUB 7337, Budker INP 96-78,
hep-ex/9610008. {\it Phys. Rev. Lett.}, {\bf 78} (1997) 4757,
erratum ibid. {\bf 80} (1998) 2747.
\vspace{-0.2cm}
\bibitem{Huang} Zh. Huang, R.D.Ruth {\it Phys. Rev. Letters}, {\bf 80}
(1998) 976.
\vspace{-0.2cm}
\bibitem{BalMikch} V.E.Balakin and A.A.Mikhailichenko, Preprint INP 79-85,
Novosibirsk, 1979.
\vspace{-0.2cm}
\bibitem{Baier97} V.N.Baier, preprint Budker INP 97-12, 1997.
\vspace{-0.2cm}
\bibitem{ChenPalmer} P.Chen, R.B. Palmer, {\it ``Advanced Accelerator
 Concepts''} proceedings: Edited by J.S. Wurtele, AIP Conference
 Proceedings, 279 (1993) 888.



%
%
\end{thebibliography}
\end{document}